\documentclass[aps,pre,twocolumn,10pt]{revtex4-1}
\usepackage{graphicx,color}
\usepackage{amsmath,amssymb}

\begin{document}
\title{Experimental phase-space-based optical amplification of scar modes}
\author{C. Michel$^1$, S. Tascu$^2$, V. Doya$^1$, P. Aschi\'eri$^1$, W. Blanc$^1$, O. Legrand$^1$ and F. Mortessagne$^1$\\}
\affiliation{$^1$Laboratoire de Physique de la Mati\`ere Condens\'ee, Universit\'e Nice-Sophia Antipolis, CNRS, UMR 7336, France\\
$^2$Research Center on Advanced Materials and Technologies, Faculty of Physics, Alexandru Ioan Cuza University of Iasi, Blvd. Carol I, nr. 11, 700506, Iasi, Romania}

\begin{abstract}
Waves billiard which are chaotic in the geometrical limit are known to support non-generic spatially localized modes called scar modes. The interaction of the scar modes with gain has been recently investigated in optics in micro-cavity lasers and vertically-cavity surface-emitting lasers. Exploiting the localization properties of scar modes in their wave analogous phase space representation, we report experimental results of scar modes selection by gain in a doped D-shaped optical fiber.
\end{abstract}
\maketitle

Wave chaos is investigating the wave behavior of bounded systems with chaotic classical dynamics. The common feature of chaotic modes in a linear (passive) medium is ergodicity \cite{Schnirelmann1974}: modes of a chaotic cavity exhibit a generic statistically uniform field distribution \cite{Berry1977}. Scar modes that result from constructive wave interferences along the least unstable classical periodic orbits break the ergodicity property \cite{Heller84} but are rather scarce events. Even though scar phenomenon has been intensively investigated since their numerical discovery by Mc Donald and Kaufman \cite{McDonald79}, their localization feature still arouses interest as in the profilic context of graphene investigations \cite{Huang09}.\\
Interaction between spatially localized scar modes and nonlinear effects is of fundamental interest and constitutes a recent evolution of Wave Chaos. The subtle interplay of gain with wave chaos and specifically scar modes has been widely examined in optics in micro-cavity lasers \cite{Gmachl1998,Fang2005,Harayama2003} and in vertically-cavity surface-emetting lasers \cite{Yu11}. The problem of the interaction between the lasing medium and the light field in a two-dimensional (2D) laser cavity is still a vivid research area \cite{Wang2009}. In the presence of a saturable nonlinear effect, which does not involve any phase change, scar modes take a crucial part on the improvement of the lasing efficiency. Thanks to their low-loss properties compared with generic ergodic modes, scar modes are commonly selected during the lasing process leading to enhanced emission directivity and lowered lasing threshold \cite{Fang2007}. In recent studies, the use of gain has also proved to favor the selection of scar modes in guided optics: numerical investigations revealed that the interaction of the chaotic modes with a localized active medium in a D-shaped optical fiber amplifier led to a selective amplification of scar modes \cite{Michel2007}.

\begin{figure}[h]
\centering
\includegraphics[width=8cm]{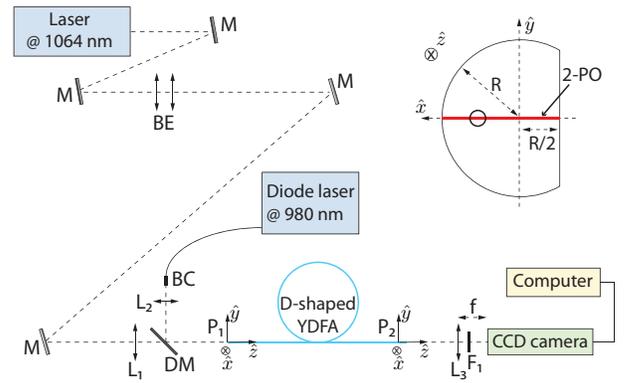}
\caption{Experimental setup for the amplification of scar modes in the double clad Ytterbium doped fiber amplifier. L$_1$, lens (15-mm working distance), L$_2$, lens (10-mm working distance), L$_3$, lens (2.5-cm focal length f), DM, dichroic mirror (T=90$\%$@1064 nm, R=90$\%$@978 nm), P$_1$, P$_2$, 3-D positioning supports, F$_1$, bandpass filter (3 nm @1064 nm), BC, beam collimator, BE, beam expander, M, mirror. Right inset, transverse cross section of the D-shaped fiber amplifier with the off-centered doped area (circle) and the 2-PO along the x-axis. The polymer cladding is not represented. Color online}
\label{Manip}
\end{figure}

In this letter, we report the first, to our knowledge, experimental demonstration of a selective amplification of scar modes through gain in a highly multimode chaotic system. This amplification is based upon the realization of a D-shaped optical fiber doped with a spatially localized gain. The amplified scar modes are built on the 2-bounce periodic orbit (2-PO) (see inset of Fig.\ref{Manip}) of the associated D-shaped billard. Using the specific features of the scar modes in their wave-analogous phase space representation and a spatially coherent amplification process, we show that the selection mechanism of scars is strong enough to enhance scar modes and make them take over ergodic modes. Indeed, we can establish that certain non-scarred modes, displaying a good spatial overlap with the gain area, remain unamplified while the neighboring scar modes are clearly amplified, essentially because those non-scarred modes are not phase-space localized in the way the 2-PO scar modes are.

Our experimental system is a non standard Ytterbium doped multimode optical fiber with a core diameter of 121 $\mu$m, truncated at its half radius (see inset in Fig.\ref{Manip}). 
\\
We fabricate the fiber preform by a standard MCVD (modified chemical vapor deposition) method. The core is made of pure silica including a 15-$\mu m$-diameter doped area of Ytterbium, free from any other co-dopant. A low refractive index polymer surrounds the silica core and acts as an optical cladding. The doped area is off-centered and located in a specific position deduced from numerical investigations \cite{Michel2007} which corresponds to the maximum of intensity for most of the 2-PO scar modes. The preform has been cut and shaped before being stretched for this purpose. The optical index mismatch between the doped area and the D-shaped core is induced by the ions concentration and takes the value $\Delta n=5\times 10^{-4}$ ($[\text{Yb}^{3+}]=900$ ppm). This value of $\Delta n$, coupled with a triangular index profile, prohibits any light guidance in the doped area. Both sides of the fiber are optically polished to avoid either at the input or at the output any diffraction induced by defects. \\
The signal is given by a YAG laser ($P_s$=400 mW) at $\lambda_s$=1064 nm and expanded thanks to a couple of microscope objectives ($\times 4$ and $\times 10$) that play the role of a beam expander (BE). The resulting beam, of 5 mm of diameter, is sent through a 15-mm working distance lens (L$_1$) and is transmitted by a dichroic mirror (DM, with 90 $\%$ of transmission at $\lambda_s$). The pump ($P_p$=10 W) is provided by a cw laser diode (coupled in an angle-polished fiber of 105-$\mu$m diameter) at $\lambda_p$=974 nm. The pump is focused thanks to a 10-mm working distance lens (L$_2$) on the input of the D-shaped optical fiber through the dichroic mirror (90 $\%$ of reflexion at $\lambda_p$). For an optimization of the pump absorption (excitation of a maximized number of modes), the beam is strongly focused into the multimode core \cite{Doya2002,Leproux2003}. \\
The pump absorption coefficient has been measured under low pump power to prevent any population inversion and takes the value $\alpha_p=6.5$ dB/m. The absorption and emission cross sections of the pump and the signal are respectively $\sigma_{pa}=2.65\times 10^{-24}$ m$^2$, $\sigma_{pe}=2.65\times 10^{-24}$ m$^2$, $\sigma_{sa}=5.56\times 10^{-26}$ m$^2$ and $\sigma_{se}=6.00\times 10^{-25}$ m$^2$. The gain of the signal has been measured along a 20 m-length fiber, and the maximum value $G= 4.5$ dB (with an output signal to noise ratio of 27 dB) is obtained at the optimal length $L_{\text{opt}}=15$ m. This gain value, which is rather low as compared to standards of telecom optical amplifiers, serves our purpose in demonstrating that amplification is really at work in our highly multimode fiber. At the output of the fiber, the residual pump is filtered by a bandpass filter $F_1$ of 3 nm-width around $\lambda_s$ and a CCD camera is placed in the focal plane of a 2.5 cm-focal length lens to image the far-field intensity pattern. As we want to observe strong signatures of waves localization associated with scar modes, we choose to vizualize the signal intensity in the far-field configuration. Indeed, the square modulus of the spatial Fourier transform of the field distribution gives some information about the directions and modulus of the outgoing wave vectors.

\begin{figure}[h!]
\centering
\includegraphics[width=8.cm]{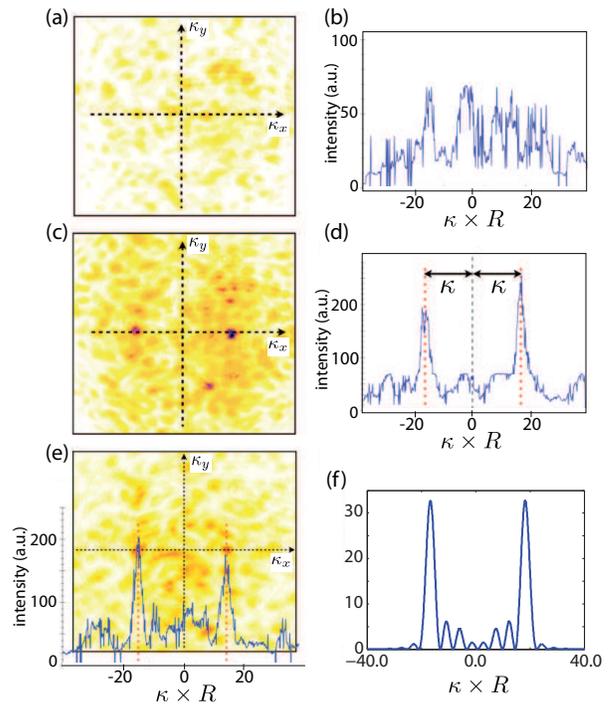}
\caption{Far field intensity observations (a), (c) and their transverse horizontal cut at $\kappa_y=0$ (b), (d) for the unpumped and pumped configuration respectively for $\kappa_p=7$. Far field intensity pattern and transverse horizontal cut superimposed for the pumped configuration with $\kappa_p=6$ (e). Horizontal cut of the far field intensity of a calculated scar mode (f). Color online}
\label{ManipResult}
\end{figure}

At the fiber entrance, we control the direction of the transverse wave vectors propagating in the fiber. A selection of the injected transverse wave vectors is obtained by a $(\hat{x},\hat{y})$-translation of the polished input of the optical fiber in the divergent part of the focused signal beam. The figure \ref{ManipResult}(a) presents the typical far field intensity of the signal obtained at the output of the multimode  15 m-length D-shaped fiber amplifier when pump is off. Even if the injection of the signal is optimized to excite a scar mode among few ergodic modes, the bending of the fiber tends to distribute the signal intensity among numerous other modes. The observed statistically uniform distribution of intensity reveals the superposition of a large number of modes in the fiber. Figure \ref{ManipResult}(b) represents a cut along the $\kappa_x$-axis for $\kappa_y=0$. It confirms the broad distribution of transverse wavenumbers.\\
For the same signal configuration, but with pump turned on, the figure \ref{ManipResult}(c) presents the far field intensity pattern of the outgoing signal. A pair of high intensity spots symmetrically distributed on both sides along the $\kappa_x$-axis appears in the center of the far field. These two peaks demonstrate that the signal propagates along the fiber with two privileged  transverse directions given by $\kappa$ and $-\kappa$ along the $x$-direction. Those directions are associated with those of the 2-PO. The observation of two symetrical peaks along the 2-PO direction is the fingerprint of a scar mode in the far-field intensity representation as can be seen in the $\kappa_x$-cut of the far-field intensity pattern of a typical calculated scar mode (Fig. \ref{ManipResult}(f)). 
\\
The transverse wavenumber corresponding to the observed experimental peaks can be estimated from figure \ref{ManipResult}(d). The calibration of the far field representation gives one pixel = (2.405$\pm$0.005)$\times 10^{-3}$ $\mu$m$^{-1}$ and then, the transverse wavenumber associated to the peaks is deduced. We get $\kappa=(0.279\pm 0.003)$ $\mu$m$^{-1}=(17.0\pm 0.2)/R$. This value has to be compared with the theoretical value deduced from a condition of phase coherence of waves that bounce back and forth along the 2-PO in the D-shaped billiard:
\begin{equation}
\kappa_p\mathcal{L}+\Delta\phi+\frac{\pi}{2}=2\pi p
\label{quantif}
\end{equation}
where $\mathcal{L}=3R$ is the 2-PO-length, $\Delta\phi$ is the phaseshift induced by the reflexion at the core/cladding interface, the additional $\pi/2$ phaseshift is a consequence of the single self-focal point of the 2-PO, and $p$ is an integer that gives the order of the scar mode. For $p=7$, we get a value $\kappa_{p=7}=17.08/R$ which is in perfect agreement with the measured value: the scar mode of order $7$ is selected thanks to the localized gain. We change the signal launching condition by translating the fiber input in the $(\hat{x},\hat{y})$ plane (Fig. \ref{Manip}) in order to seed the amplification process on another scar mode. Figure \ref{ManipResult}(e) reports the enhancement of two symetrical peaks along the $\kappa_x$-direction for this new input illumination. The measured value of the transverse wavenumber $\kappa=(15.0\pm 0.2)/R$ has to be compared to the calculated value (Eq. \ref{quantif}) $\kappa_{p=6}=15.0/R$. The agreement is again remarkable and one can deduce that the scar mode of order $p=6$ is enhanced thanks to amplification. These two examples are the proof that, by controlling the initial illumination condition, one can select a given scar mode along the 2-PO thanks to the optical amplification process.

\begin{figure}[h!]
\centering
\includegraphics[width=8.1cm]{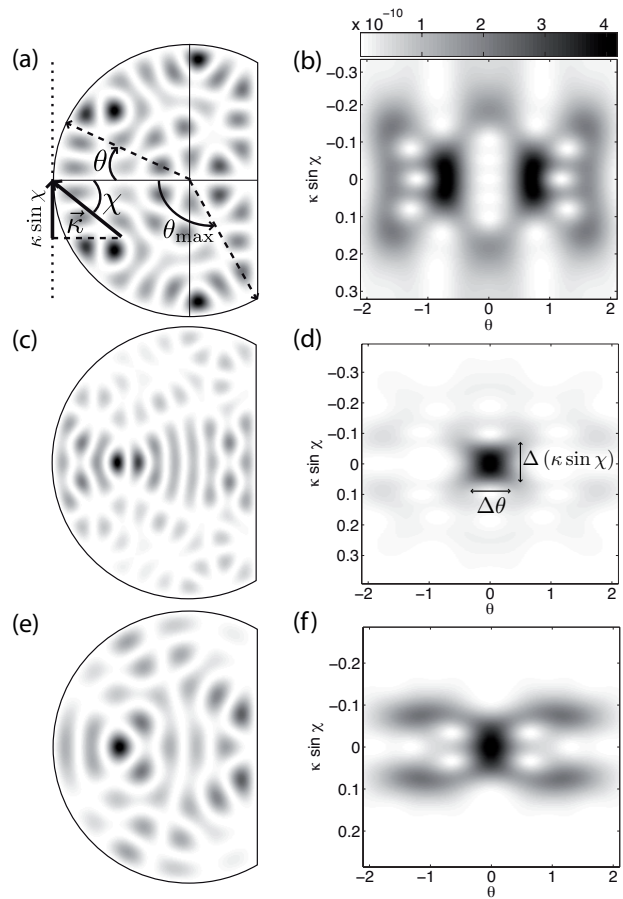}
\caption{Field intensity of specific modes and their respective Husimi representations. For a generic speckle mode (a),(b); for  a scar mode along the 2-PO (c), (d); and for the mode with $\kappa_{\mathrm{NS}}$ (e),(f). Color online}
\label{Husimi}
\end{figure}

These preliminary experimental results demonstrate our ability to selectively amplify scar modes in a highly multimode fiber thanks to a localized gain. In agreement with common fiber amplification processes, the spatial overlap of the modes with the active area is usually the main control parameter of the amplification efficiency.
We propose a complementary interpretation of the scar modes enhancement by gain based on the physical nature of scars. According to E. J. Heller, ``scars manifest themselves as enhanced probability in phase space, as measured by overlap with coherent states placed on the periodic orbits'' \cite{Heller84}. Therefore, scar modes, viewed as fingerprints of classical orbits in the near-field intensity pattern, also show strongly privileged wave directions as highlighted through their far-field intensity. Both spatial and directional features of scar modes can be most efficiently combined in the so-called Husimi representation \cite{Husimi1940}. The Husimi representation is commonly used as a wave equivalent to the classical Poincar\'e surface of section. Thus, it establishes 
a strong correspondence between some particular modes and their associated classical trajectories in the semiclassical regime. In the case of 2D cavities, the Husimi function measures the normal derivative of the eigenfunction on the boundaries of the cavity  \cite{Birkhoff1930,Crespi1993,Ree1999,Backer2004}. Fig. \ref{Husimi}(b-d-f) present the Husimi representation of three different calculated modes of the chaotic optical fiber with metallic boundary in the 2-dimensional space ($\theta,\kappa\sin\chi)$ with $\theta$ the curved abscissa, $\kappa$ the transverse wavenumber of the modes and $\kappa\sin\chi$ the projection of the transverse wavevector along the tangential direction to the boundary (Fig. \ref{Husimi}(a)).
The Husimi representation (Fig. \ref{Husimi}(b)) of a generic ergodic mode (Fig. \ref{Husimi}(a)) generally explores a large part of the available space: no specific transverse wavevector directions appear to be prevailing. This observation is consistent with the hypothesis of generic modes built upon an ergodic semiclassical behavior  \cite{Berry1977,Berry1981}. Ergodic modes result from the superposition of a large number of plane waves with fixed transverse wavenumber $\kappa$ but random directions $\chi$. On the contrary, the Husimi representation (Fig. \ref{Husimi}(d)) of a scar mode associated to the 2-PO (Fig. \ref{Husimi}(c)) exhibits a strong localization of the field around the angular abscissa $\theta_0=0$ and the exclusive direction of the 2-PO, that is, $\chi_0=0$. All the 2-PO scar modes present such a similar localization attribute in their Husimi representation. 
\\
The spatial and directional extensions of the Husimi pattern define a range of angular abscissa $\Delta\theta$ around $\theta_0$ and of transverse wavevector directions $\Delta\chi$ derived from $\Delta(\kappa\sin\chi)$ around $\chi_0$. Thus, a scar mode is built upon few highly directional plane waves, ensuring a strong spatial coherence. In a geometrical approach in terms of rays in the D-shaped billiard, this span of initial conditions extracted from the Husimi distribution determines an area in the billiard plane where rays are converging. This area is located in the vicinity of the so-called self-focal point of the 2-PO. This location of the active ions guarantees an efficient round-trip of rays through the active medium, as well as a spatially coherent amplification and also a good overlap with the field amplitude of the scar modes. We stress that all these conditions should be fulfilled to assure a scar mode amplification.
Indeed, we observe that non-scarred (NS) modes can be insensitive to amplification despite a good spatial overlap with the gain medium. For instance, the field amplitude of the NS mode of Fig.\ref{Husimi}(e) has a good spatial overlap with the doped area but is not selected during the amplification process \cite{note}. Using a Beam Propagation Method algorithm \cite{Feit78}, we simulate the propagation of a gaussian-like initial signal injected into the fiber with a transverse wavenumber $\kappa_{\mathrm{NS}}=17.07/R$ that corresponds to the specific NS mode. The modulus of the transverse wavevector is fixed, but its direction is chosen so as to avoid the 2-PO direction and thus a 2-PO scar mode excitation. 
\begin{figure}[h!]
\centering
\includegraphics[width=8.1cm]{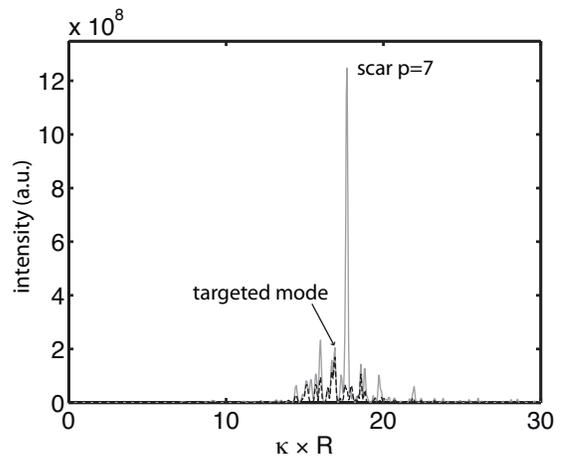}
\caption{Intensity spectrum at the input of the fiber amplifier (dashed) and after amplification at the output of the fiber (gray) associated to a gaussian signal launched into the fiber with $\kappa_{\mathrm{NS}}$. Color online }
\label{Spectre51}
\end{figure}\\
Fig.\ref{Spectre51} presents the intensity spectrum at the output of the D-shaped fiber amplifier. The targeted NS mode is not amplified whereas the scar mode of order $p=7$ is strongly amplified despite its reduced contribution in the initial spectrum: the spatial field overlap is not able to clarify this favored amplification of the scar mode. The explanation is supplied in the analysis of the Husimi representation (Fig. \ref{Husimi}(f)). Indeed, the NS mode presents a broad distribution of transverse wavevectors directions in the Husimi representation that prohibits a coherent amplification mechanism. As a consequence, a strong localization in the Husimi representation around the signature of the 2-PO is essential to ensure the efficiency of modes amplification. 

In this paper, we have presented an experimental observation of scar modes amplification in the D-shaped optical fiber with localized gain.  An active medium with a spatial location controlled by the phase space signature of scars is a propitious system to selectively amplify scar modes. Investigations about scar modes in nonlinear systems are opening new attractive issues. For instance, having a scar amplifier may offer new perspectives for optical communications. The intensity spectrum of a scar modes amplifier has a response similar to the spectral response of a Fabry-Perot cavity. Each scar mode appears well-separated with a spacing between adjacent transverse wavenumbers equal to $2\pi/\mathcal{L'}$ with $\kappa\mathcal{L'}=2\pi p$. These well-resolved modes injected in a multimode fiber would be less sensitive to crosstalk and could be employed as independent transmission channels in a process of mode division multiplexing \cite{Koebele2011}. 
\\
We express grateful thanks to S. Trzesien and M. Ude for their priceless involvement in the preform and fiber manufacture. 

\end{document}